\magnification=\magstep1 
\baselineskip 18 truept 
\voffset 2 truecm
\tolerance=20000 
\hskip 5truecm
\vskip 1 truecm\noindent 
\centerline{\bf A Percolative Model of Soft Breakdown in Ultrathin Oxides} 
\medskip\noindent 
\centerline{$^1$C.  Pennetta, $^1$L. Reggiani and $^2$Gy.  Tref\'an} 
\medskip 
\centerline{$^1$INFM - National
Nanotechnology Laboratory, Dipartimento di Ingegneria dell'Innovazione}
\centerline{Universit\`a di Lecce, Via Arnesano, I-73100, Italy}
\centerline{$^2$Eindhoven University of Technology, Dept.  of Electrical 
Engineering}
\centerline{5600 MB Eindhoven, The Netherlands} 
\vskip 0.5 truecm\noindent 
\centerline{\bf Abstract} 
\vskip 0.5 truecm\noindent 
The degradation of ultrathin oxide layers in the presence of a 
stress voltage is modeled in terms of two antagonist percolation 
processes taking place in a random resistor network.  
The resistance and leakage current fluctuations are studied by Monte Carlo
simulations for voltages below the breakdown threshold.
An increase of excess noise together with a  noticeable 
non-Gaussian behavior is found in the pre-breakdown
regime in agreement with experimental results.
\vfill\eject
%
%
{\bf 1. Introduction and model} 
\vskip1pc\noindent
Soft breakdown occurs in
ultrathin oxide layers during constant voltage stress.
This process manifests itself by a large increase of the stress-induced leakage
currents (SILC) and by the occurrence of giant current fluctuations
associated with resistance fluctuations.
These fluctuations can be related to trapping and detrapping of
electrons in the percolation cluster formed between the electrodes of the
capacitor at the soft breakdown [1,2].
In this paper, we investigate the resistance in the pre-breakdown region
and the associated fluctuations using two
percolations which evolve in competition on a random resistor network [3], as
briefly summarized below.
We describe a thin film as a two-dimensional
square-lattice network of resistors of resistance $r_n$, laying on an
insulating substrate at temperature $T_0$ acting as a thermal reservoir.
Geometrical effects arising from different shapes are neglected. 
Initially all the resistors are identical, $r_n=r_0$.
We take a square geometry $N \times N$ where $N$ determines
the linear sizes of the lattice and $N_{tot} = 2 N^2$ is the total number of
resistors.
Electrical contacts, realized by perfectly conducting bars,
are placed at the left and
right hand sides of the network where a constant voltage $V$ is applied.
By increasing the applied voltage, defects are introduced by replacing
single resistors with short circuits. Thus, the network degrades by undergoing an
insulator-conductor transition when the defects gradually become so dense
that finally they form a continuous short circuit path between the contacts
[3].
The temperature $T_n$ of the $n$-th resistor is computed by [3]:
$$
T_{n}=T_{0} + A \Bigl[ r_{n} i_{n}^{2} + {B \over N_{neig}}
\sum_{l=1}^{N_{neig}}  \Bigl( r_{l} i_{l}^2   - r_n i_n^2 \Bigr) \Bigr]
\eqno(1)
$$
where the term which adds to $T_0$ in the r.h.s. of Eq. (1) accounts for
Joule heating associated with the current flowing in the resistor and the
coupling with the nearest neighbours.
Here, $N_{neig}$ is the number of first neighbours around the $n$-th
resistor with current $i_n$, $A$ the
heat coupling parameter and $B=3/4$ provides a uniform heating
everywhere in the perfect network.
We note that Eq. (1) assumes an instantaneous thermalisation
between neighbor resistors.
Probabilities of creating and recovering a defect are taken as
$W_D=exp(-E_D/K_BT_n)$ and $W_R=exp(-E_R/K_BT_n)$ with
$E_D$ and $E_R$ characteristic  activation energies, respectively, and $K_B$
the Boltzmann constant.
Monte Carlo simulations are carried out using the following procedure.
(i) Starting
from the perfect lattice, we calculate the total network resistance $R$,
the external current $I$, and the
local currents by solving Kirchhoff's loop equations by the Gauss elimination
method.  
The local temperatures are then calculated and used for the successive
update of the network.  (ii) The defects are generated with probability
$W_D$.
The local currents and the local temperatures are then recalculated.
(iii) The
defects are recovered with probability $W_R$, and the total network
resistance, the
local currents, and the temperatures are finally calculated.  This procedure is
iterated from (ii).  Each iteration step can be associated with an elementary
time step.  The iteration runs until either the defect percolation threshold is
reached (i.e.  the network resistance drops below $10^{-3}$ 
times to its initial value), or steady-state is achieved.  
In this latter case, the iteration is let to
run long enough for a fluctuation analysis to be carried out.  
\vskip1pc\noindent
{\bf 2.  Results and discussion} 
\vskip1pc\noindent
We compare the results of numerical simulations to the corresponding
experimental observations and validate the present approach by showing that
it reproduces the key experimental features. Since soft breakdown experiments 
report leakage current evolutions, noise spectra and probability density of
fluctuations,  we investigate the same quantities by numerical simulations.
In the simulations we applied a constant voltage stress with the following
realistic parameters $N = 75$, $r_0 = 10^7 \ \Omega$, 
$T_0=300 \ K$, $A=5 \times 10^7 \ ^oC/W$, $E_D = 0.19 \  eV$, 
$E_R = 0.13 \ eV$. Figure 1 reports the leakage current evolutions for stress 
voltages, respectively, of $0.1$, $0.3$, $0.5$, $0.6$, $0.7$, $0.8$, $0.85$ 
and $0.9 \ V$. We note that the values of the activation energies $E_D$ and 
$E_R$ control the speed of the evolution toward steady state and/or breakdown 
and as such should be calibrated from a  quantitative comparison with 
experiments. However, the qualitative features of the breakdown process we are
presently interested in do not depend significantly from their absolute values.
We found that for voltages lower than $0.9 \ V$  no breakdown occurs but a 
dynamical balance takes place between defect generation and defect recovery.
The steady state so achieved exhibits current fluctuations whose statistical 
and spectral features characterize the electrical quality of the dielectric.
However, for stress voltages equal or larger than $0.9 \ V$ after a few 
thousand iteration steps the leakage current grows more than a thousand fold 
which indicates a breakdown situation. We also observe that the breakdown does
not occur as a result of a systematic current growth but more as a result of 
current bursts. In fact, a set of bursts generally occurs that lead to the 
breakdown of the insulator.  
Thus an interesting pre-breakdown region appears like in experiments [4,5].
From a detailed analysis of a set of simulations we also conclude that 
a threshold voltage exists below which no breakdown occurs.
Figure 2 shows how the breakdown voltage depends on the
resistance of the virgin dielectric, modeled as an initial perfect lattice
with increasing value of the elemental resistor.  This breakdown voltage is
found to increase proportional to the square root of the perfect
network  resistance.
Figure 3 shows the steady state current noise spectrum calculated by FFT for
stress voltages ranging from $0.1$ to $0.85 \ V$.
Within numerical uncertainity, we detect Lorentzian
spectra of the same corner frequency which indicates that the characteristic
times of fluctuations are independent of stress voltages.
Interestingly, the strong increase of the value of
the plateaux at increasing  stress voltages shows a
super-quadratic increase of the noise with the applied stress.
The current-noise spectra in Fig. 3 are found to be in qualitative agreement
with experiments [4].
Figure 4 reports the normalized variance of resistance
fluctuations as a function of the applied voltage.
Starting from the intrinsic value of the network [6], the variance is
found to increase significantly as a net effect of the stressing voltage
which is ultimately responsible of the breakdown at about $0.9 \ V$.
Recent observations of SILC measurements reported non-Gaussian current 
fluctuations in the soft breakdown region of ultra-thin dielectrics [1].  
Therefore, we computed the distribution
function of the current fluctuations $P(I)$ for stress voltages
of $0.5 \ V$ and  $0.85 \ V$ and plotted them in Fig. 5 as a function of 
$I \ - <I>$, where $<I>$ is the average current value.
On this linear-log plot the corresponding
Gaussian distribution would be a downward parabola.
The distribution for an applied voltage near breakdown of $0.85 \ V$ 
clearly exhibits a non-Gaussian tail at 
high currents over the average due to the large upward
bursts which are not balanced by downward bursts in the current evolution.
Since a further increase of the stress voltage causes the film to break down,
the emerging non-Gaussian distribution can be thought of as a precursor of
failure as observed in experiments [1].  
\vskip1pc\noindent
{\bf 3.  Conclusions} 
\vskip1pc\noindent
We have developed a
percolative approach to study the soft breakdown and the associated current
fluctuations of ultrathin insulating films.The major results we found are
summarized as follows.
(i) The current evolution exhibits a pre-breakdown region.
(ii) A threshold voltage depending on the resistance of the otherwise perfect
film separates breakdown and steady state conditions.  
(iii)
The spectrum of the current fluctuations is basically of Lorentzian type with a
correlation time which is practically independent of the stress voltage.  
(iv) The variance of resistance fluctuations exhibits a giant  
super-quadratic enhancement at increasing the stress voltage.  
(v) At high enough voltage stress the current fluctuations 
exhibit a non-Gaussian behavior.  
These features, particularly the super-quadratic
and the non-Gaussian behaviors of the current noise at high stress, are of
relevant interest as failure precursors.  These results are in satisfactory
qualitative agreement with experimental findings.  
\vskip1pc\noindent
\centerline{\bf Acknowledgements} 
\vskip1pc\noindent
This research is performed within the STATE project of INFM. 
 Partial support is also provided by CNR MADESS II project. 
\medskip\noindent 
\centerline{\bf References} 

\item{[1]}
N. Vandewalle et al.,  Appl.  Phys.  Lett.  74 (1999) 1579.

\item{[2]}
M.  Houssa et al., Appl.  
Phys.  Lett.  73 (1998) 514.

\item{[3]}
C. Pennetta et al.,  IEEE Trans.  on Electron.  Dev.  47 (2000) 1986.

\item{[4]}
G. Alers et al., Appl.  Phys.  Lett.  69 (1996) 2885.

\item{[5]}
D.J. DiMaria and J.H. Stathis, Appl.  Phys.  Lett.  70 (1997) 2708

\item{[6]} C.  Pennetta et al., Phys.  Rev. Lett.  85, (2000) 5238.

\vfill\eject \centerline{\bf Figure Captions}
\vskip 1truecm\noindent
Figure 1 -  Leakage current evolutions in a degrading dielectric kept under
constant voltage for stress voltages (from bottom to top) of:
$0.1$, $0.3$, $0.5$, $0.6$, $0.7$, $0.8$, $0.85$ and $0.9 \ V$.

\vskip 1truecm\noindent
Figure 2 - Breakdown voltage $V_b$ as a function of the perfect
network resistance $R_{per}$. The graph shows $V_b \propto \sqrt{R_{per}}$.

\vskip 1truecm\noindent
Figure 3 -  Power spectral density of current fluctuations of the
steady-state dielectric film kept under constant voltage of
(from bottom to top): $0.1$, $0.3$, $0.5$, $0.6$, $0.7$ and $0.85 \ V$.

\vskip 1truecm\noindent
Figure 4 -  Normalized variance of the network resistance fluctuations 
as a function of the applied voltage.

\vskip 1truecm\noindent
Figure 5 -  Gaussianity check of current fluctuations for a stress voltage
of $0.5 \ V$ (open squares) and of $0.85 \ V$ (filled circles).

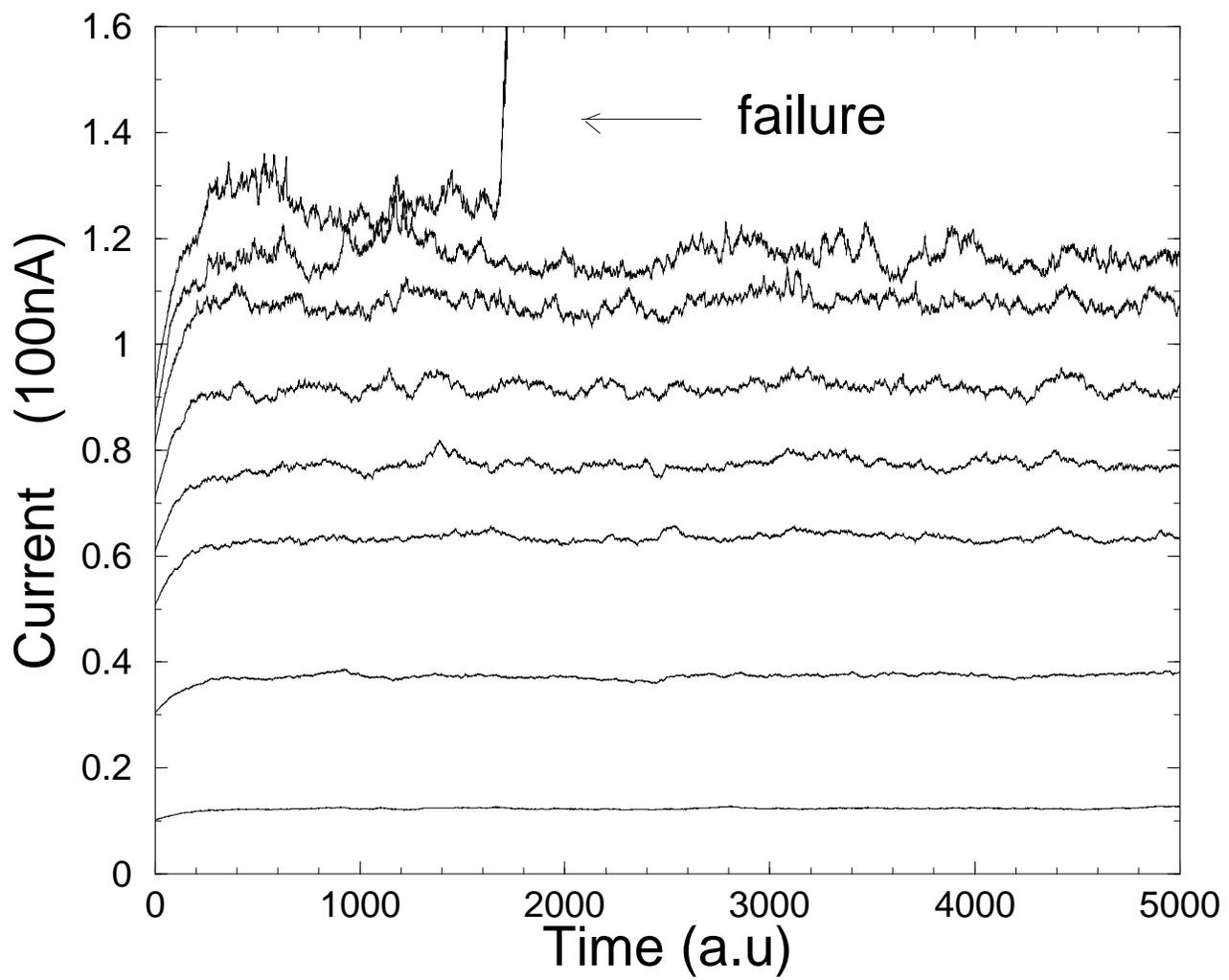
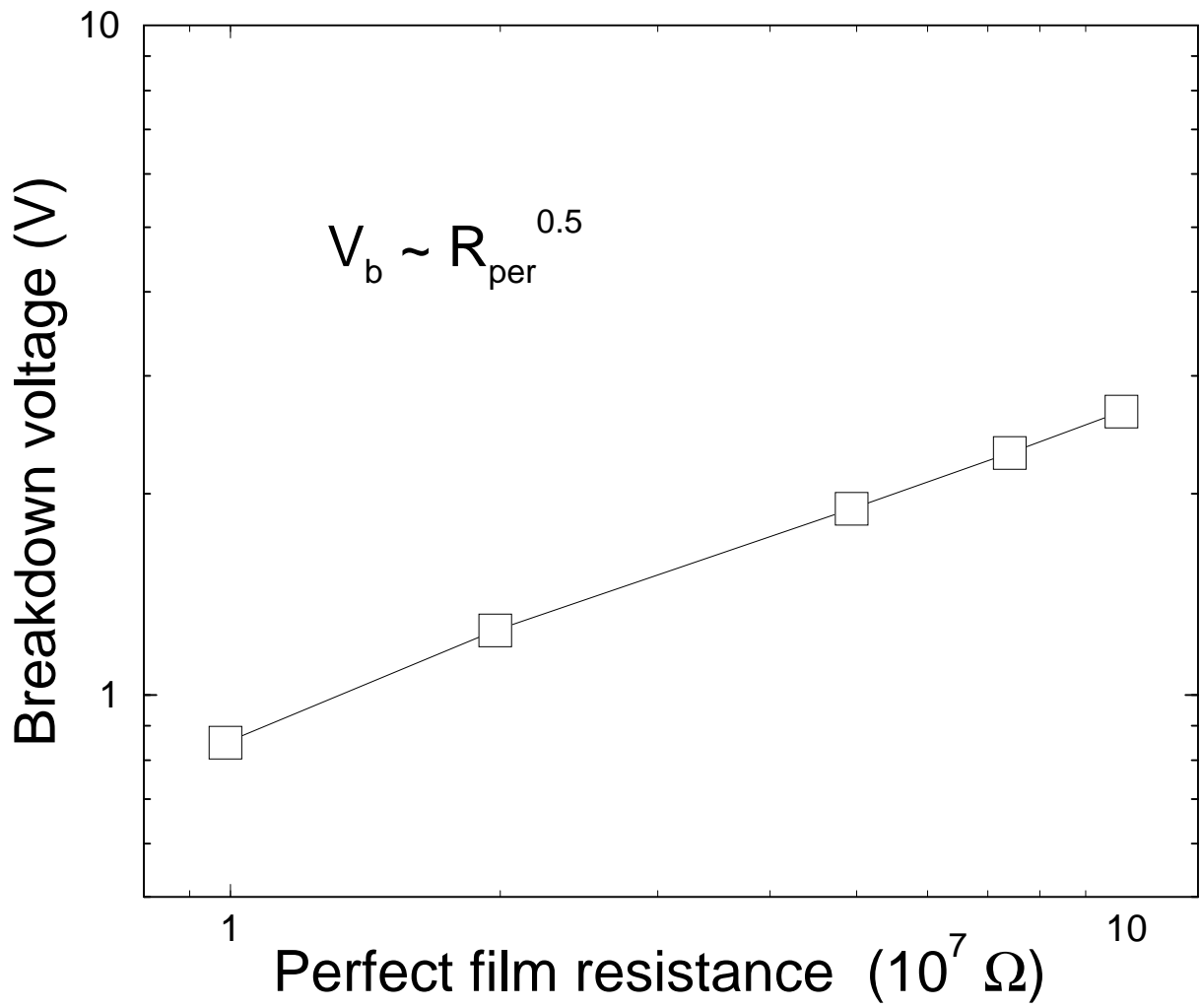
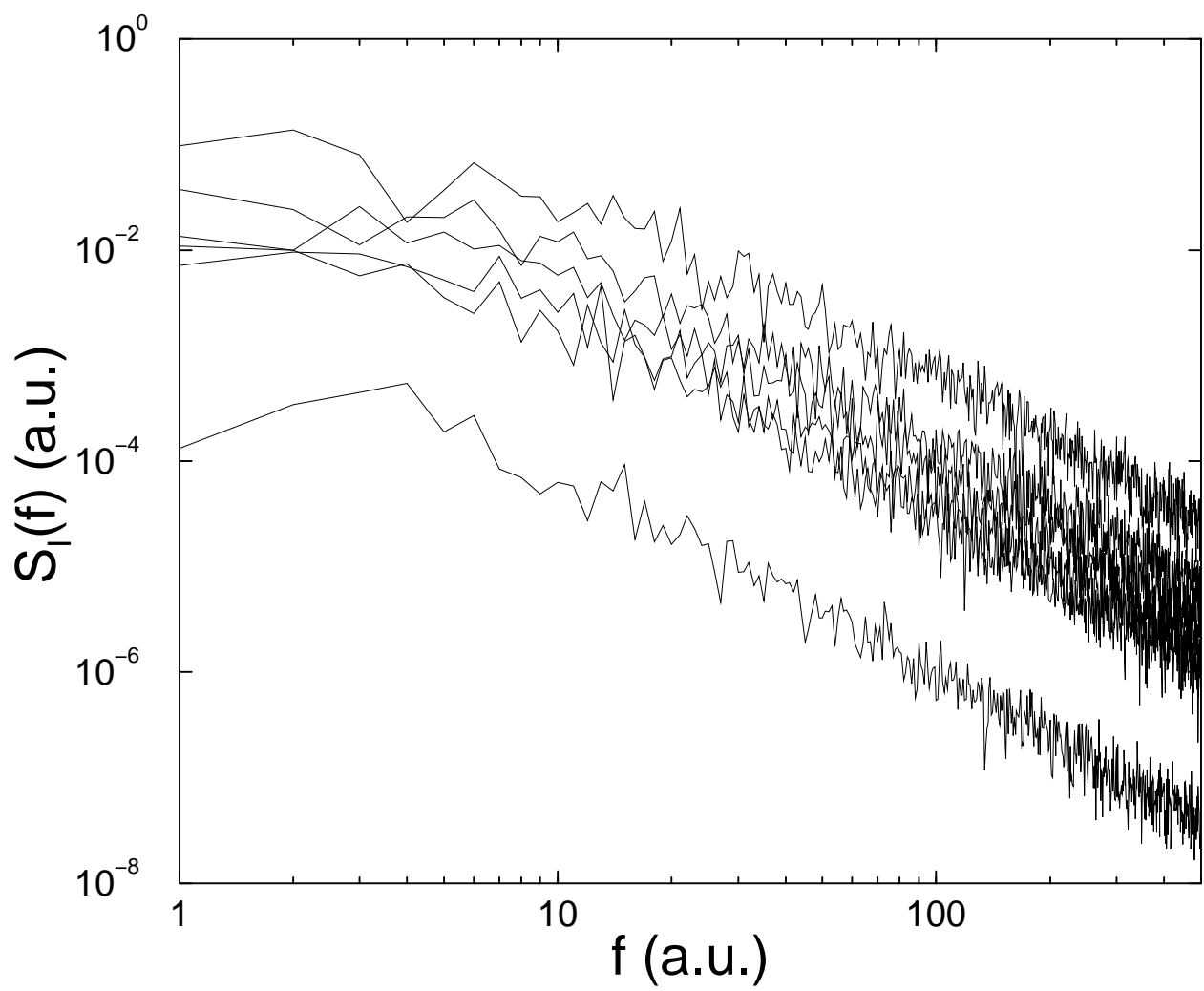
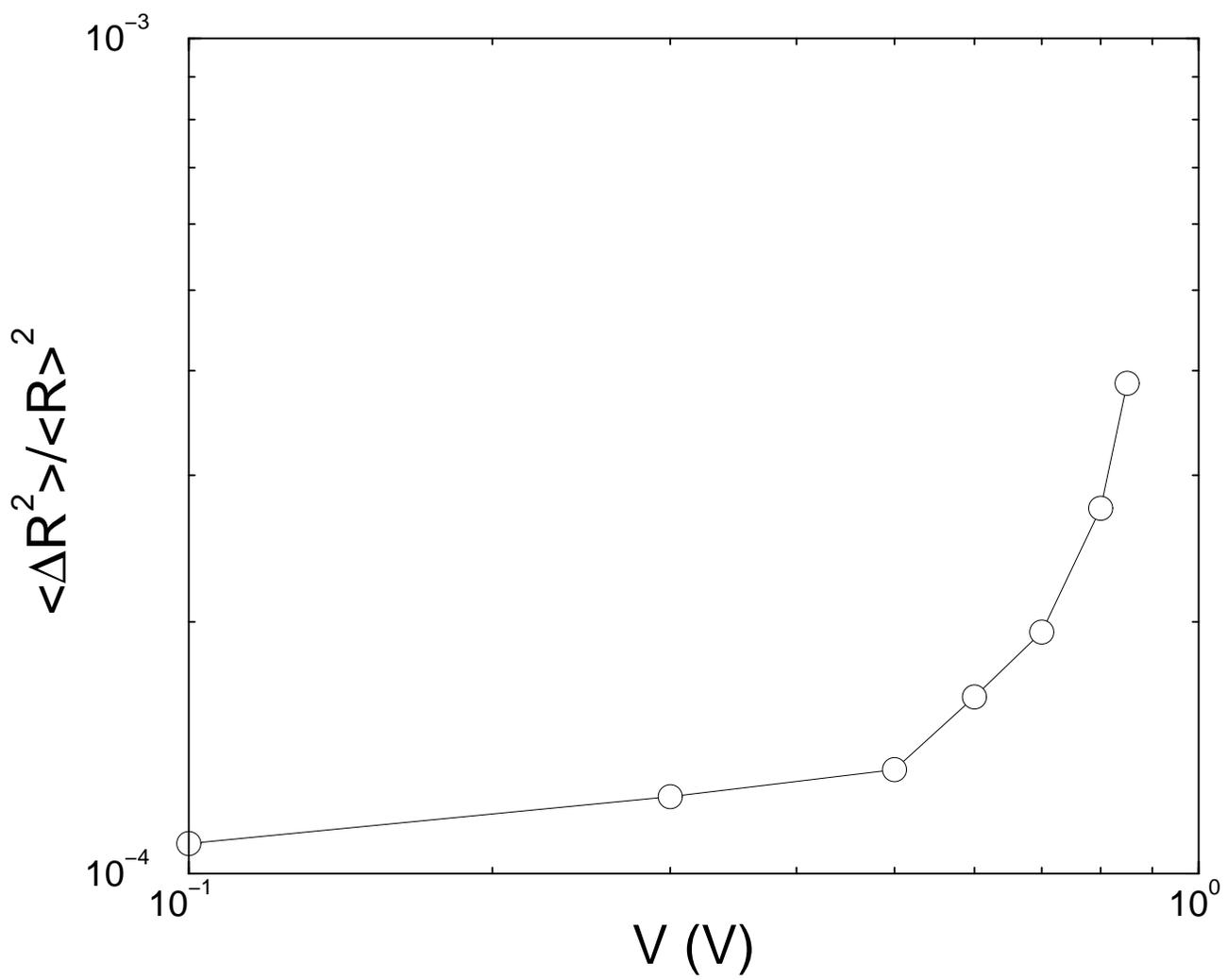
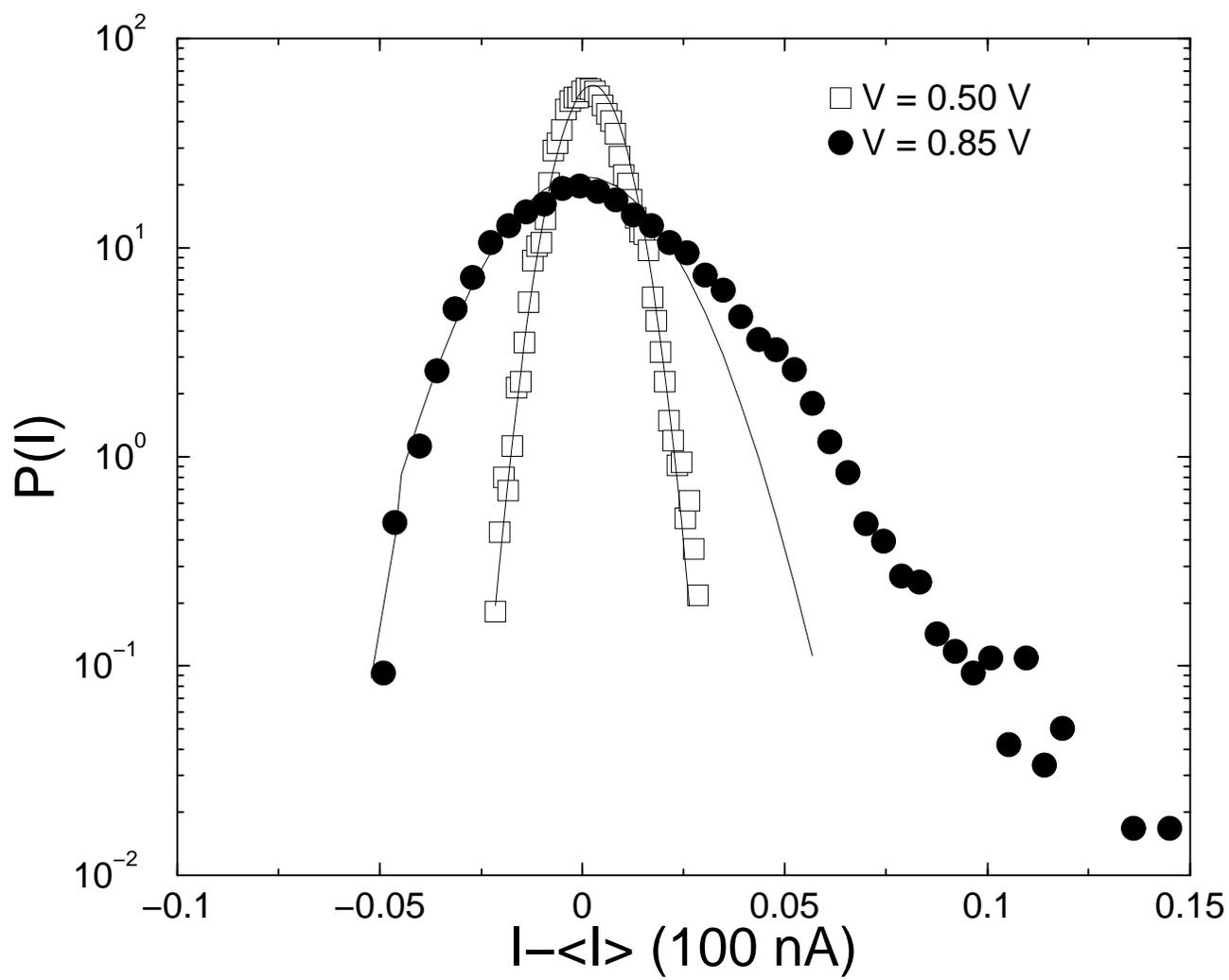
\bye